\documentclass[aps,pre,preprint,showpacs,groupedaddress]{revtex4}
\usepackage[dvips]{graphicx}
\begin{document}

\title{Effective delivering capacity in traffic dynamics based on scale-free network}

\author{Shan He}
\author{Sheng Li}
\email{lisheng@sjtu.edu.cn}
\author{Hongru Ma}
\affiliation{Department of Physics, Shanghai Jiao Tong University,
Shanghai 200240, China}


\begin{abstract}
We investigate the percentage of delivering capacities that are
actually consumed in a typical traffic dynamics where the
capacities are uniformly assigned over a scale-free network.
Theoretical analysis, as well as simulations, reveal that there
are a large number of idle nodes under both free and weak
congested state of the network. It is worth noting that there is a
critical value of effective betweenness to classify nodes in the
congested state, below which the node has a constant queue size
but above which the queue size increases with time. We also show
that the consumption ratio of delivering capacities can be boosted
to nearly $100\%$ by adopting a proper distribution of the
capacities, which at the same time enhances the network efficiency
to the maximum for the current routing strategy.
\end{abstract}


\maketitle

\section{Introduction}
The interplay between the dynamics and the topology of network can
be used to model the evolutions and structures of a wide variety
of complex systems in nature and human society. A branch of
dynamics research that focuses on the data traffic in the network
has drawn a lot of interests. The studies in this field are mainly
about capturing the attributes of the data transportation and
providing hints on how to make the physical communication network
more efficient. Previous structure analysis based on real data
spanning several disciplines revealed that many networks,
including the Internet, exhibit scale-free
characteristics~\cite{albert:47}. The degree distribution of these
networks follows a power law. Having explored the topological
properties~\cite{newman:016132,goh:278701}, researchers habitually
choose the scale-free network as the underlying structure when
studying the traffic dynamics in order to observe phenomena that
are close to those in the real
world~\cite{pastor-satorras:3200,kim:027103,adamic:046135}.

The traffic models proposed by recent works generally consist of
four components: the underlying network topology, the information
packet generation, the routing strategy and the packet delivering
capacity of individual
nodes~\cite{zhao:026125,guimera:248701,chen:036107,ohira:193,wang:026111}.
They all have counterparts in the reality. Take the Internet as an
example. The four components correspond to the physical links
between computers, the average data flow, the algorithm implemented
on routers and the number of parallel packet processors per site,
respectively. They also have other interpretations in the field of
path navigation~\cite{arenas:124} and organization
design~\cite{radner:1109,decanio:275}. In these models, a continuous
phase transition of network from free state to congested state is
observed when the packet generation rate exceeds a critical
value~\cite{ohira:193}. The core purpose of researches on these
models is the optimization that enhances the network capacity which
is measured by the critical value of packet generation rate. Some of
the previous efforts concentrate on finding a most suitable network
topology~\cite{chen:036107,guimera:248701,zhao:026125}, while the
others are mainly about deciding an optimal routing
strategy~\cite{adamic:046135,kim:027103,echenique:056105,chen:036107}.
However, in most of the models, constant delivering capacity is
assigned to nodes. It is not economical as the loads on different
nodes vary. In fact, the delivering capacities are processor
resources that they should also be taken into consideration during
optimization attempts.

With the hope of discovering potential possibilities of improvements
on the communication network, we analyze the percentage of
delivering capacities that are truly active in our traffic model
where the capacities are uniformly distributed over a scale free
network. It is concerned that congestion is sometimes inevitable in
real lives, for example, everyone has the experience of traffic jam
in rush hours, so the performance of delivering processors under
congested state of network is paid special attentions. We find that
some of the nodes in a weak jammed network always have constant
number of packets waiting to be delivered which is the same
situation as in the free state of the network, whereas the others
have increasing number of packets. These two types can be
distinguished by referring to the effective betweenness which is an
attribute of node defined in Ref.~\cite{guimera:248701}. Our
theoretical estimates and simulations reveal that delivering
capabilities of nodes are not well utilized even under the congested
state of network, though optimal routing strategy based on local
degree information discussed in Ref.~\cite{wang:026111} is adopted.
To alleviate the waste of processor resources, we apply a
non-uniform distribution of the delivering capacities which adjusts
the network to the best performance under the given conditions.

This paper is organized as follows. The traffic model is described
in Sec. II. In Sec. III, theoretical analysis and simulations of
the model are provided in both free and congested state. The
conclusion is given in Sec. IV.

\section{Traffic Model}
We use the famous model proposed by Barab\'{a}si and
Albert~\cite{barabasi1999esr} to build the underlying scale-free
network. BA model, which features growth and preferential
attachment, uncovers for the first time the mechanism controlling
the emergence of power-law degree distribution observed in real
networks. We set the model parameters $m_{0}=m=5$ and network size
$N=1000$. In the traffic dynamics, network nodes are thought to be
sources that produce information packets, routers that deliver
packets and first-in-first-out queues with unlimited size that store
to-be-processed packets. At each time step, for a node
$i(i=1,\ldots,N)$, the following procedures are done. A packet is
generated with probability $\rho_{i}$. The new packet whose
destination is randomly selected among the other $N-1$ nodes is
placed at the back of queue $i$. Meanwhile, packets are picked from
the front of queue $i$. If the picked packet has destination node
$i$, it is removed; otherwise, it is delivered to node $j$, one of
the neighbors of node $i$, with preferential weight
\begin{equation}\label{eq_rs}
\Pi_{j}=\frac{k^{-1}_{j}}{\sum_{l}k^{-1}_{l}},
\end{equation}
where sum runs over the neighbors of node $i$ and $k_{i}$ is the
degree. In Eq.~(\ref{eq_rs}), we have used the optimal routing
strategy based on the knowledge of only neighbor's
degree~\cite{wang:026111}. At most $C_{i}$ packets can be processed
by node $i$ if there are adequate ones in queue $i$. $C_{i}$ is the
delivering capacity of node $i$. We set $\rho_{i}=\rho$,
$C_{i}=C=10$ following the parameters used in
Ref.~\cite{wang:026111} in order to meet the conditions for the
optimal routing strategy. These procedures are carried out for
different $i$ in parallel. What we concern is the ratio between the
number of packets actually processed by the network during a unit
time and the total capacities assigned, which is defined as the
effective delivering capacity.

\section{Effective delivering capacity}
Depending on $\rho$, there is a continuous phase transition from
free state to congested state~\cite{arenas:124}. When $\rho$ is
small, the packets flow freely in the network as the nodes always
have available processing abilities to send them to the next
positions. The average travel time $\tau$ of the packets keeps
constant and the total number $N_{p}(t)$ of packets floating in the
network at time step $t$ fluctuates slightly around $\rho N\tau$.
However, when $\rho$ is large, packets beyond the delivering
capacities accumulate continuously in the queues, causing $\tau$ to
diverge. Since the number of packets the network can manage and the
generation rate $\rho$ are both constants, the amount of
accumulation at every time step is also constant. As a result,
$N_{p}(t)$ increases linearly with time. To watch the transition
accurately, we use the order parameter $\xi$ introduced in
Ref.~\cite{arenas:3196},
\begin{equation}
\xi=\lim_{t\rightarrow\infty}\frac{1}{\rho N}\frac{\langle
N_{p}(t+\Delta t)-N_{p}(t)\rangle}{\Delta t},
\end{equation}
where $\langle\ldots\rangle$ indicates an average over time windows
of width $\Delta t$. From the properties of $N_{p}(t)$, we know that
$\xi$ is zero in free state but non-zero in congested state.
$\rho_{c}$, the critical value of $\rho$, which is also the maximum
generation rate keeping the system in free state, measures the
network capacity.

To characterize the situation in free state, we use the method
introduced in Ref.~\cite{guimera:248701} with our modifications. Let
us focus on a packet at node $i$ whose destination is node $k$. The
probability for this packet to go to node $j$ in one time step is
denoted as $p_{ij}^{k}$. The precise form of $p_{ij}^{k}$ depends on
the routing strategy. For the scheme used in the model,
\begin{equation}
p_{ij}^{k} =
(1-\delta_{ik})\frac{A_{ij}k_{j}^{-1}}{\sum_{l}A_{il}k_{l}^{-1}},
\end{equation}
where $A_{ij}$ is the element of the adjacent matrix. The strategy
is Markovian that each packet is delivered independently. As the
network is in free state, packets pass through nodes without any
wait. The probability for the above packet to go to node $j$ in $n$
time steps is given by
\begin{equation}
P_{ij}^{k}(n)=\sum_{l_{1},l_{2},\ldots,l_{n-1}}p_{il_{1}}^{k}p_{l_{1}l_{2}}^{k}\cdots
p_{l_{n-1}j}^{k}.
\end{equation}
Treating $p_{ij}^{k}$ and $P_{ij}^{k}(n)$ as elements of matrices
$\mathbf{p}^{k}$ and $\mathbf{P}^{k}(n)$ respectively, we have
\begin{equation}
\mathbf{P}^{k}(n)=(\mathbf{p}^{k})^{n}.
\end{equation}
Let us switch the focus to the centrality of node. In the traffic
model, packets coming from neighbors at a specific time step are a
fraction of packets generated before. Supposing that one packet with
target node $k$ is generated at node $i$ each time step, we
calculate $b_{ij}^{k}(t)$, the average number of such packets moving
to node $j$ at time step $t$,
\begin{equation}\label{eq_bkt}
\mathbf{b}^{k}(t)=\sum_{n=1}^{t}(\mathbf{p}^{k})^{n}.
\end{equation}
As $N_{p}(t)$ is stable, when $t\rightarrow\infty$,
\begin{equation}\label{eq_bk}
\mathbf{b}^{k}=\sum_{n=1}^{\infty}(\mathbf{p}^{k})^{n}=(\mathbf{I}-\mathbf{p}^{k})^{-1}\mathbf{p}^{k}.
\end{equation}
Summing over all the possible sources and targets of packets yields
the effective betweenness of node $j$, $B_{j}$,
\begin{equation}
B_{j}=\sum_{i,k}b_{ij}^{k}.
\end{equation}
Note that $B_{j}$ depends on both the routing strategy and the
network topology. As the routing algorithm can vary, highly
connected nodes do not always have large effective betweenness. The
load $L_{j}$, which is the average number of packets to be delivered
by node $j$ every time step, is written as,
\begin{equation}\label{eq_l}
L_{j}=\rho+\frac{\rho B_{j}}{N-1},
\end{equation}
where the first term on the right side corresponds to packets
generated by $j$ itself while the second term corresponds to packets
coming from neighbors. Any node $j$ with $L_{j}>C$ will cause the
network to be congested. Thus we estimate $\rho_{c}$ by
\begin{equation}\label{eq_rhoc}
\rho_{c}=\frac{C}{\frac{B^{*}}{N-1}+1}\approx \frac{C(N-1)}{B^{*}},
\end{equation}
where $B^{*}$ is the maximum effective betweenness. In our model, we
had $\rho_{c}=0.0045$. When $\rho<\rho_{c}$, the number of packets
processed at node $j$ per time step is exactly its load $L_{j}$.
Therefore, effective delivering capacity $\eta$ under free state is
\begin{equation}\label{eq_edcfree}
\eta=\frac{\sum_{j}L_{j}}{CN},
\end{equation}
which is proportional to $\rho$.

For $\rho$ larger than $\rho_{c}$, Eq.~(\ref{eq_edcfree}) is invalid
because the system is not expected to be in free state when
$t\rightarrow\infty$ in Eq.~(\ref{eq_bk}). Back to time step $t=1$,
the average load in the network is equal to the generation rate
$\rho$. Here we limit our discussion to a weak congested scenario
where $\rho$ is not supposed to be extremely large so that the nodes
are still able to process all the packets in the queues for the
first few time steps. During these time steps, we can obtain the
instantaneous load $L_{j}(t)$ by replacing Eq.~(\ref{eq_bk}) with
Eq.~(\ref{eq_bkt}) in the previous calculation for free state. As
$L_{j}(t)$ is a monotonically increasing function of time, it is at
a later time step $t_{c}$ that the largest $L_{j}(t)$ starts to
exceed the delivering capacity $C$. While the packets beyond the
capacity are hindered at the corresponding node, they do not flow in
the network. Comparing with an imaginary case where no constraints
on the delivering capacities are applied, the actual instantaneous
load increases with time more slowly. Though $L_{j}(t)$ for
$t>t_{c}$ can not be directly worked out by analytical calculation,
we can predict that with more nodes suffering from packet
overloading, $L_{j}(t)$ increases even more slowly. After a
transient time, the instantaneous load over the whole network stops
increasing and remains stationary. Thereafter, all the nodes in the
network maintain their states. Those with load higher than
delivering capacity have growing queues, while the others keep
constant queues. Since the effective betweenness $B_{j}$ determines
free state load $L_{j}$ by Eq.~(\ref{eq_l}), it is plausible to
suppose that nodes with greater effective betweenness become
overloaded more easily. We expect a critical effective betweenness
$B_{c}$ for current $\rho$, which can be used as the criterion to
classify the two types of node. Once $B_{c}$ is found, effective
delivering capacity can be considered by separate examinations on
different types of node.

\begin{figure}[tb]
\includegraphics[height=65mm]{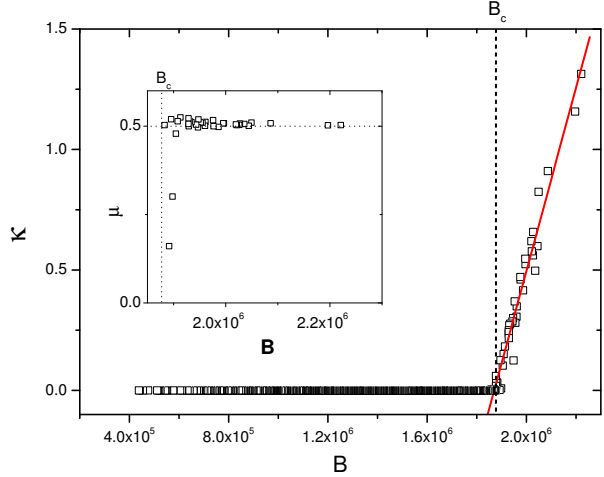}
\caption{\label{grapha}Simulation results of $\kappa$ versus $B$.
The critical point $B_{c}$ is found by calculation. In the inset, we
plot $\mu_{j}=(Q_{j}(2t)-Q_{j}(t))/Q_{j}(2t)$ when
$t\rightarrow\infty$ for nodes with $B_{j}>B_{c}$, which shows that
the queues increase linearly with time.}
\end{figure}

To find $B_{c}$, we look at the congested state of system when
$t\rightarrow\infty$. The increasing queues are already considerably
long. A newly hindered packet at a specific node will wait in the
queue for so long a time that it seems to be disappeared from the
views of the other nodes in the network. We assume that a node with
an increasing queue is a ``target'' for information transportation
where packets disappear. To make it more like a ``target'', the $C$
packets it lets out at every time step are shifted to the neighbors
as equivalent generation rates. By this imagination, the original
congested network turns out to be a smaller one in free state, and
method for $\rho<\rho_{c}$ can be exploited. We use the following
algorithm:
\begin{enumerate}
\item All the nodes in the network are sorted by effective
betweenness $B_{j}$ in descending order. The first node in the
sequence is added to an initially empty set $S$.
\item\label{item_two} The new generation rate $\rho'_{j}$ for node $j$ is
\begin{equation}
\rho'_{j}=\rho+\sum_{i\in S}Cp_{ij},
\end{equation}
where sum runs over all the elements in $S$ and $p_{ij}$ represents
the routing strategy,
\begin{equation}
p_{ij}=\frac{A_{ij}k_{j}^{-1}}{\sum_{l}A_{il}k_{l}^{-1}}.
\end{equation}
\item In order to suppress packets arriving at ``targets'',
row $i$ in the original matrix $\mathbf{p}^{k}$ is set to 0 for each
$i\in S$.
\item Employing $\rho'_{i}$ and the modified $\mathbf{p}^{k}$, the new load
$L'_{j}$ is evaluated by
\begin{equation}
L'_{j}=\left\{\begin{array}{ll} 0 & j\in S\\
\rho+\frac{1}{N-1}\sum_{i,k}\rho_{i}'b_{ij}^{k} & \textrm{otherwise}
\end{array}\right.
\end{equation}
\item\label{item_five} The validity of
\begin{equation}\label{eq_equbc}
max(L_{j}')<C
\end{equation}
must be ensured, which corresponds to constant queue size for nodes
not in $S$. If there exists load $L'_{j}$ larger than $C$, the next
node in the sequence descending ordered by $B_{j}$ is added to $S$.
\item Step \ref{item_two} to \ref{item_five} are repeated until
Eq.~(\ref{eq_equbc}) is fulfilled. $B_{c}$ for current $\rho$ is
just the critical $B_{j}$ separating nodes within $S$ from those
outside.
\end{enumerate}
To describe the changes of queues in the simulation, we define
$\kappa_{j}$ for node $j$,
\begin{equation}
\kappa_{j}=\lim_{t\rightarrow\infty}\frac{\langle Q_{j}(t+\Delta
t)-Q_{j}(t)\rangle}{\Delta t},
\end{equation}
where $Q_{j}(t)$ is the queue size of node $j$ at time step $t$ and
$\langle\ldots\rangle$ indicates an average over time windows of
width $\Delta t$. $\kappa_{j}$ is zero if node $j$ has a constant
queue or positive if node $j$ has an increasing queue. Taking
$\rho=0.02$ as an example, we illustrate $\kappa_{j}$ versus $B_{j}$
in Fig.~\ref{grapha}, where $B_{c}$ is obtained by calculation. It
is shown that below $B_{c}$ nodes have constant queues but above
$B_{c}$ the queues increase linearly with time. Besides,
$\kappa_{j}\sim B_{j}-B_{c}$ for $B_{j}>B_{c}$.

We are ready to estimate the effective delivering capacity under
congested state. For the nodes with constant queues, the packets
delivered are equal to the load $L'_{j}$. For the nodes with
increasing queues, they work at the capacity $C$. Note that during
the calculation, packets processed by these nodes are shifted to the
neighbors, which indicates that the packets pass through nodes
twice. The effective delivering capacity for congested state,
$\eta'$, is then given as
\begin{equation}\label{eq_edccong}
\eta'=\frac{\sum_{j}L_{j}'+2CN_{s}}{CN},
\end{equation}
where $N_{s}$ is the number of elements in $S$.

\begin{figure}[tb]
\includegraphics[height=65mm]{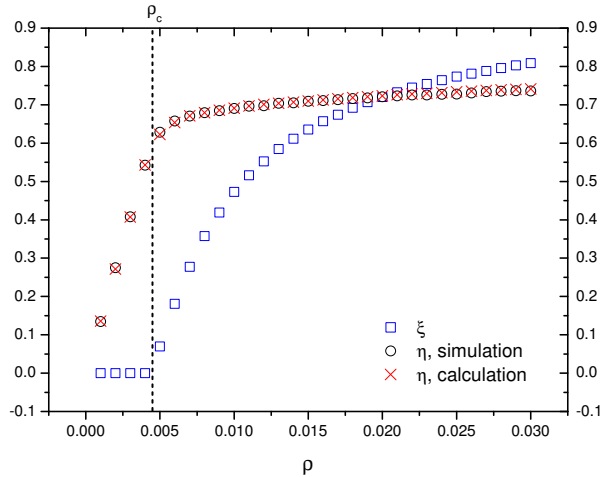}
\caption{\label{graphb}Effective delivering capacity $\eta$ and
order parameter $\xi$ for $C=10$. $\rho_{c}$ follows
Eq.~(\ref{eq_rhoc}).}
\end{figure}

As shown in Fig.~\ref{graphb}, we calculated the effective
delivering capacity for different $\rho$ ranging from $0.001$ to
$0.03$ and compared the results with simulation where packets
processed in a unit time are directly counted. In the free state,
$\eta$ increases with $\rho$ according to Eq.~(\ref{eq_edcfree}). In
the congested state, though the increment of $\rho$ shows a trend to
enlarge $L'_{j}$, there is a cutoff imposed by the delivering
capacity $C$. With the overloaded nodes ignored, the sum of $L'_{j}$
does not vary too much. $\eta'$ increases only due to the increment
of $N_{s}$ which is small when $\rho$ changes hardly.

\begin{figure}[tb]
\includegraphics[height=65mm]{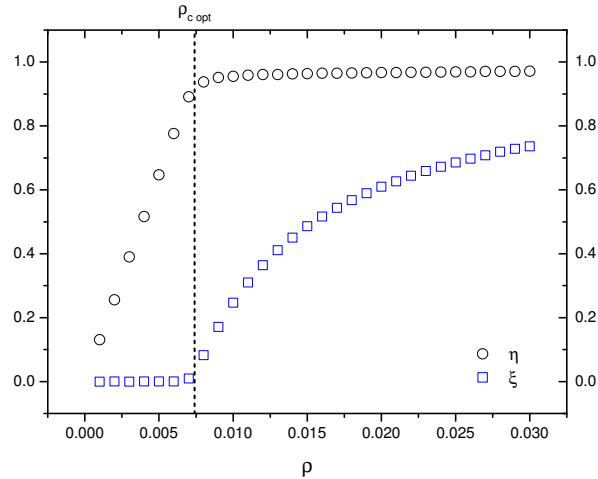}
\caption{\label{graphc}Effective delivering capacity $\eta$ and
order parameter $\xi$ for new distribution of $C$ captured by
simulation. $\rho_{c~opt}$ follows Eq.~(\ref{eq_rhocopt}).}
\end{figure}

It is undesirable that up to $\rho=0.03\approx6.7\rho_{c}$ the
effective delivering capacity is no more than $75\%$, despite the
use of optimal local routing algorithm. In other words, while the
network is not able to handle all the information packets, $25\%$
processors are idle. An optimization problem arises naturally: given
the fixed number of processors, how to exploit them all? The key is
to avoid resource wastes on less loaded nodes. We introduce a new
delivering capacity distribution,
\begin{equation}\label{eq_copt}
C_{i}'=\frac{B_{i}}{\sum_{j}B_{j}}CN,
\end{equation}
where the total capacity $CN$ is conserved. With respect to
Eq.~(\ref{eq_l}), when $\rho$ is approaching the critical value, the
delivering capacity just meets the load, and no waste happens. Above
the critical value of $\rho$, cutoff imposed by delivering capacity
does not reduce the load on the nodes with constant queues, and the
utilization ratio keeps $100\%$. In particular, the network capacity
is enhanced to the maximum since all the processors are active.
Similar to Eq.~(\ref{eq_rhoc}), for an arbitrary distribution of
delivering capacity, the critical point $\rho_{c}$ is
\begin{equation}\label{eq_rhocarb}
\rho_{c}=min\{\frac{C_{j}(N-1)}{B_{j}},j=1,\ldots,N\},
\end{equation}
with constraint $\sum_{j}C_{j}=CN$. The optimal value of $\rho_{c}$
is obtained by neutralizing the differences among $C_{j}(N-1)/B_{j}$
under the constraint. It can be accomplished by employing
Eq.~(\ref{eq_copt}). The corresponding critical point $\rho_{c~opt}$
is
\begin{equation}\label{eq_rhocopt}
\rho_{c~opt}=\frac{CN(N-1)}{\sum_{j}B_{j}}.
\end{equation}
Calculation yields $\rho_{c~opt}=0.0074$. In the simulation,
delivering capacity given by Eq.~(\ref{eq_copt}) is converted to
integer where errors are involved. Figure~\ref{graphc} illustrates
simulation results of the new distribution. It is clear that the
effective delivering capacity is very close to $100\%$ for
$\rho>\rho_{c~opt}$ and $\rho_{c~opt}$ is consistent with
theoretical estimates.

\section{Conclusion}
In this paper, we study the effective delivering capacity for a
typical traffic model under both free and weak congested state. It
is shown that with equal number of packet processors assigned to
nodes, a large percentage of the processors are idle even under the
congested state. By redistributing the processors according to
effective betweenness, all of them are activated, which corresponds
to maximum network capacity for the routing strategy adopted.
Considering the difficulties to change the topologies of real
networks, previous works~\cite{wang:026111,chen:036107} on
optimization mainly focus on finding the routing algorithm best fits
the underlying structure. We rise the importance of choosing a
proper distribution of the delivering capacity which is an essential
supplement to the optimized routing strategies. In addition, we
successfully separate nodes with constant queues from those with
increasing queues in a weak congested network by referring to a
critical value of effective betweenness, which may provide some
hints for further studies on the congested network.

\begin{acknowledgements}
This paper was supported by the National Science Foundation of
China under Grant No. 10105007 and No. 10334020.
\end{acknowledgements}

\end{document}